\documentclass[prl,twocolumn,aps,superscriptaddress,showpacs]{revtex4}
\usepackage{epsfig}
\def\beq{\begin{equation}}
\def\eeq{\end{equation}}
\def\bea{\begin{eqnarray}}
\def\eea{\end{eqnarray}}
\def\<{\langle}
\def\>{\rangle}

\usepackage{graphicx}
\usepackage[dvips]{color}
\usepackage{colordvi}
\usepackage{amsfonts}
\usepackage{amssymb}
\usepackage{amsmath}
\usepackage{graphics}
\usepackage{pstricks}
\usepackage{bm}
\usepackage{natbib}
\begin{document}                          
%%%%%%%%%%%%%%%%%%%%%%%%%%%%%%%%%%%%%%%%%%%%%%%%%%%%%%%%%%%%%%%%%%%%%%%%%
\title{Shear viscosity of neutron matter from realistic nucleon-nucleon interactions}
%%%%%%%%%%%%%%%%%%%%%%%%%%%%%%%%%%%%%%%%%%%%%%%%%%%%%%%%%%%%%%%%%%%%%%%%%
\author{Omar Benhar}
\affiliation{INFN, Sezione di Roma. I-00185 Roma, Italy}
\affiliation{Dipartimento di Fisica, Universit\`a ``La Sapienza''. 
I-00185 Roma, Italy}
\author{Marco Valli}
\affiliation{Dipartimento di Fisica, Universit\`a ``La Sapienza''.
I-00185 Roma, Italy}
\affiliation{INFN, Sezione di Roma. I-00185 Roma, Italy}
%%%%%%%%%%%%%%%%%%%%%%%%%%%%%%%%%%%%%%%%%%%%%%%%%%%%%%%%%%%%%%%%%%%%%%%%%
%%%%%%%%%%%%%%%%%%5
\date{\today}
%\keywords{yyy}
%%%%%%%%%%%%%%%%%%%
\begin{abstract}
The calculation of transport properties of Fermi liquids, based on
the formalism developed by Abrikosov and Khalatnikov, requires the knowledge 
of the probability of collisions between quasiparticles in the vicinity of 
the Fermi surface.
We have carried out a numerical study of the shear viscosity of pure
neutron matter, whose value plays a pivotal role in determining the stability 
of rotating neutron stars, in which these processes are described
using a state-of-the-art nucleon-nucleon potential model. Within our approach 
medium modifications of the scattering cross section are consistently 
taken into account, 
through an effective interaction obtained from the matrix elements of the
bare interaction between correlated states. Inclusion of medium effects 
leads to a large increase of the viscosity at densities larger 
than $\sim 0.1$ fm$^{-3}$.
\end{abstract}
\maketitle
%%%%%%%%%%%%%%%%%%%%%%%%%%%%%%%%%%%%%%%%%%%%%%%%%%%%%%%%%%%%%%%%%%%%%%%%%
Viscosity plays a pivotal role in determiming the stability of rotating neutron stars.
As Chandrasekhar first pointed out \cite{chandra}, emission of gravitational radiation
(GR) following the excitation of non-radial oscillation modes may lead to the instability
of rotating stars.
While this effect would make all perfect fluid rotating stars unstable, in presence
of viscosity dissipative effects damp the oscillations, and may prevent the onset of
the instability.
As a consequence, a quantitative understanding of the viscosity of neutron star matter 
is required
to determine whether a mode is stable or unstable
(for a recent review on neutron star oscillations and instabilities see, e.g., 
Ref. \cite{lind_1} and references therein).

%For a given viscosity, only stars rotating with angular velocity lower than a critical 
%value $\Omega_c$ are stable. Obvioulsly, as viscosity is temperature dependent, so 
%is $\Omega_c$. Foe example, the 

%It has been shown %\cite{lind1}
%that the GR driven instability associated with the fundamental $f$-mode cannot
%significantly reduce the spin of a neutron star and that substantial amounts
%f GR cannot be emitted this way; later on it was noticed that the $r$-modes
%oscillations of rotating stars whose restoring force is the Coriolis force)
%ere also subject to the GR instability.
%oreover, the GR instability is also strong enough to overcome internal
%issipation processes in neutron star matter, even in relatively slowly
%otating stars. It has been pointed out that, as the GR instability in
%r$-modes appears capable of significantly reducing the angular momentum of a
%otating star, the GR emitted during such spin-down events may well be
%etectable by LIGO %\cite{ligo}

%In order to determine which stars are stable, a detailed analysis of their
%perturbations must be carried out including the influence of both GR
%\textit{and} viscosity. The existing results %\cite{cls} 
%suggest that viscosity do not completely suppress the instability.
%%%%%%%%%%%%%%%%%%%%%%%%%%%%%%%%%%%%%%%%%%%%%%%%%%%%%%%%%%%%%%%%%%%%%%%%%%%%%%%%%%%%%%
Early estimates of the shear viscosity coefficients of neutron star 
matter were obtained in the 70s by Flowers and Itoh, who 
used the measured scattering phase shifts to estimate the neutron-neutron 
scattering probability \cite{flowi1,flowi2}. Based on the these results, 
Cutler and Lindblom carried out a systematic study of the effect of the viscosity
on neutron star oscillations, using a variety of different models of equation of 
state (EOS) of neutron star matter \cite{cut_lind}.  

The procedure followed by the authors of Ref. \cite{cut_lind}, while allowing for 
a quantitative analysis of the damping of neutron star oscillations, cannot be regarded as 
fully consistent. Ideally, the calculation of transport properties of neutron star matter 
and the determination of its EOS should be carried out 
using the {\em same} dynamical model. The work discussed in this paper 
is aimed at making a first step towards this goal.
%can be seen as a first step in this direction. 

We have computed the shear viscosity 
of pure neutron matter using a realistic nucleon-nucleon (NN) potential, the 
Argonne $v_{18}$ model \cite{av18}, previously employed to obtain the 
state-of-the-art EOS of Akmal, Pandharipande and Ravenhall \cite{APR}. 
Within our approach, based on the formalism of Correlated-Basis-Function (CBF) 
perturbation theory \cite{CBF1,CBF2}, medium modifications of the 
NN scattering cross section are also consistently taken into account, through an 
effective interaction derived from the same NN potential. 

The theoretical description of transport properties of normal Fermi liquids
is based on Landau theory \cite{baym-pethick}.
Working within this framework and including the leading 
term in the low-temperature expansion, Abrikosov and Khalatnikov \cite{ak} obtained
the approximate expression of the shear viscosity coefficient 
\beq
\label{eta_AK}
\eta_{AK} = \frac{1}{5}\rho m^\star v^2_F \tau \,\frac{2}{\pi^2(1-\lambda_\eta)} \ ,
\eeq
where $\rho$ is the density, $v_F$ is the Fermi velocity and $m^\star$ and $\tau$ denote 
the quasiparticle effective mass and lifetime, respectively. The latter can be 
written in terms of the angle-averaged scattering probability $\langle W \rangle$ 
according to 
\beq
\label{tau_AK}
\tau T^2 = \frac{8\pi^4}{{m^*}^3}\ \frac{1}{\langle W \rangle} \ ,
\eeq
with
\beq
\label{Wavg}
\langle W \rangle = \int \frac{d\Omega}{2\pi}\ \frac{W(\theta,\phi)}{\cos{\theta/2}} \ .
\eeq
Note that the scattering process involves quasiparticles on the Fermi surface. 
As a consequence, for any given density $\rho$, the scattering probability only depends on 
the angular variables $\theta$ and $\phi$, the magnitude of all quasiparticle momenta 
being equal to the Fermi momentum $p_F=(3\pi^2 \rho)^{1/3}$.
Finally, the quantity $\lambda_\eta$ appearing in Eq.(\ref{eta_AK}) is defined as
\beq
\lambda_\eta = \frac{\langle W ( 1-3\sin^4{\theta/2}\sin^2{\phi}) \rangle}
{\langle W \rangle} \ .
\eeq
The exact solution of the equation derived in Ref. \cite{ak}, obtained by 
Brooker and Sykes \cite{sb1,sb2}, reads
\bea
\nonumber
\eta & = & \eta_{AK} \ \frac{1-\lambda_\eta}{4} \\
 & \times &  \sum_{k=0}^\infty \frac{4k+3}{(k+1)(2k+1)[(k+1)(2k+1)-\lambda_\eta]} \ , 
\label{eta_sb}
\eea
the size of the correction with respect to the result of Eq.(\ref{eta_AK}) being 
$0.750 < (\eta/\eta_{AK}) < 0.925$.

Eqs.(\ref{eta_AK})-(\ref{eta_sb}) show that the key element in the determination 
of the viscosity 
is the in-medium NN scattering cross section. In Ref. \cite{panpiep}, the relation 
between NN scattering in vacuum and in nuclear matter has been analyzed under
the assumption that the nuclear medium mainly affects the flux of incoming 
particles and the phase space available to the final state particles, while leaving
the transition probability unchanged. Within this picture $W(\theta,\phi)$ can
be extracted from the NN scattering cross section measured 
in free space, $(d\sigma/d\Omega)_{\rm{vac}}$, according to
\beq
\label{Wfree}
W(\theta,\phi) = \frac{16 \pi^2}{{m^\star}^2}
\left( \frac{d\sigma}{d\Omega} \right)_{{\rm vac}}  \,
\eeq
where $m^\star$ is the nucleon effective mass and $\theta$ and $\phi$ are related to 
the kinematical variables in the center of mass frame
through $E_{cm} = p_F^2(1-\cos \theta)/(2m)$, $\theta_{cm} = \phi$. 

The above procedure has been followed in Ref. \cite{haensel}, whose authors 
have used the available tables of vacuum cross sections obtained from partial wave 
analysis \cite{SAID}. 
In order to compare with the results of Ref. \cite{haensel}, 
we have first carried out a calculation of the viscosity using Eqs.(\ref{eta_AK})-(\ref{Wfree})
and the free space neutron-neutron cross section obtained from the Argonne $v_{18}$ potential
\beq
v_{ij}=\sum_{n=1}^{18} v_{n}(r_{ij}) O^{n}_{ij} \ .
\label{av18:1}
\eeq
In the above equation
\beq
O^{n \leq 6}_{ij} = [1, (\bm{\sigma}_{i}\cdot\bm{\sigma}_{j}), S_{ij}]
\otimes[1,(\bm{\tau}_{i}\cdot\bm{\tau}_{j})] 
\label{av18:2}
\eeq
where $\bm{\sigma}_{i}$ and $\bm{\tau}_{i}$ are Pauli matrices acting in 
spin and isospin space, respectively,
and 
\beq
S_{ij}=\frac{3}{r_{ij}^2}
(\bm{\sigma}_{i}\cdot{\bf r}_{ij}) (\bm{\sigma}_{j}\cdot{\bf r}_{ij})
 - (\bm{\sigma}_{i}\cdot\bm{\sigma}_{j}) \ .
\eeq
The operators corresponding to $p=7,\ldots,14$ are associated with the 
non static components of the NN interaction, while those
 corresponding  to $p=15,\ldots,18$ account for charge symmetry violations.
Being fit to the full Nijmegen phase shifts data base, as well as to
low energy scattering parameters and deuteron properties, the Argonne $v_{18}$ potential
provides an accurate description of the measured cross sections by construction.

\begin{figure}[htb]
%\vspace*{1.in}
\centerline
{\epsfig{figure=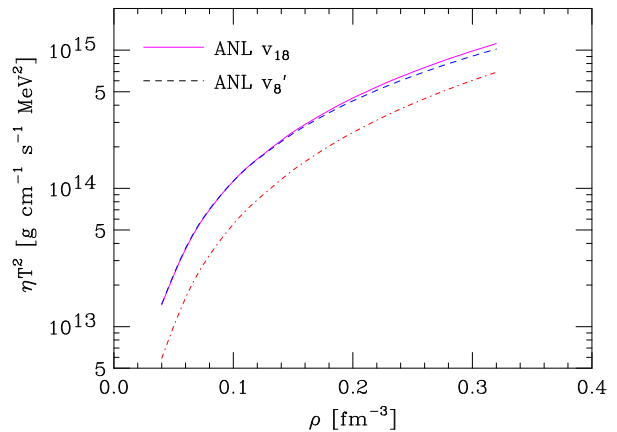,angle=000,width=8.0cm}}
%\vspace*{.2in}
%\vspace*{.2in}
\caption{ (Color online)
Neutron matter $\eta T^2$ as a function of density. Solid line: results obtained from 
Eqs.(\ref{eta_AK})-(\ref{Wfree}) using the Argonne $v_{18}$ potential and 
$m^\star$ computed from to the effective interaction described in the text. 
Dot-dash line: results obtained from Eqs.(43) and (46) of Ref. \cite{haensel} 
using the same $m^\star$. Dashed line: same as the solid line, but with the 
Argonne $v_{18}$ replaced by its reduced form $v^\prime_{8}$.
\label{etaT2_1} }
\end{figure}

In Fig. \ref{etaT2_1}, we show the quantity $\eta T^2$ as a function of density. 
Our results are represented by the solid line, while the dot-dash line corresponds 
to the results obtained from Eqs.(43) and (46) of Ref. \cite{haensel} 
using the same effective masses, computed from the effective interaction discussed below. 
The differences between the two curves are likely to be ascribed to the 
correction factor of Eq.(\ref{eta_sb}), not taken into acount by the authors of 
Ref. \cite{haensel}, and to the extrapolation needed to determine the cross sections 
at small angles within their approach. 

To gauge the model dependence of our results, we have replaced the full Argonne $v_{18}$ 
potential with its simplified form, referred to as $v^\prime_{8}$ \cite{V8P}, which only 
includes the six static operators of Eq.(\ref{av18:2}) and the two spin-orbit operators 
${{\bf L}\cdot {\bf S}}\otimes[1,(\bm{\tau}_{i}\cdot\bm{\tau}_{j})]$. 
This eight operators are the minimal set required to describe NN scattering in $S$ and $P$
states.
The corresponding results, represented by the dashed line, show that using the $v^\prime_{8}$
potential leads to a few percent change of $\eta T^2$ over the density range corresponding to 
$1/4 < (\rho/\rho_0) < 2$, $\rho_0 = 0.16$ fm$^{-3}$ being the equilibrium density 
of symmetric nuclear matter.

%%%%%%%%%%%%%%%%%%%%%%%%%%%%%%%%%%%%%%%%%%%%%%%%%%%%%%%%%%%%%%%%%%%%%%%%%%%%%%%%%%%%%
%%%%%%%%%%%%%%%%%%%%%%%%%%%%%%%%%%%%%%%%%%%%%%%%%%%%%%%%%%%%%%%%%%%%%%%%%%%%%%%%%%%%%

To improve upon the approximation of Eq.(\ref{Wfree}) and 
include the effects of medium-modifications of the NN scattering amplitude, we
have replaced the bare NN potential with an {\em effective interaction}, derived within 
the CBF approach as discussed in Ref. \cite{shannon}. 

The {\em correlated} states of neutron 
matter are obtained from the Fermi gas (FG) states through the transformation
\begin{equation}
|n\rangle = F |n_{FG}\rangle \ ,
\end{equation}
where the operator $F$, embodying the correlation structure induced by the NN 
interaction, is written in the form
\begin{equation}
F=\mathcal{S}\prod_{ij} f_{ij} \  ,
\end{equation}
$\mathcal{S}$ being the symmetrization operator. The two-body correlation functions $f_{ij}$, 
whose operatorial structure reflects the complexity of the NN potential, can be written
in the form 
\beq
f_{ij}=\sum_{n=1}^6 f^{n}(r_{ij}) O^{n}_{ij} \ ,
\label{def:corrf}
\eeq
with the $O^n_{ij}$ given by Eq.(\ref{av18:2}).

The effective interaction $v{_{\rm eff}}$ is defined by the relation
\beq
\label{def:veff}
\frac{\langle n | H | n \rangle}{\langle n | n \rangle} = 
\langle n_{FG} | T + v_{{\rm eff}}| n_{FG} \rangle  \ ,
\eeq
where $H$ is the full nuclear hamiltonian and $T$ is the kinetic 
energy operator. 
Realistic models of $H$ include, in addition to the NN potential $v_{ij}$, a three-nucleon 
potential $V_{ijk}$ needed to account for the measured binding energies of the few-nucleon 
systems, as well as the empirical equilibrium properties of symmetric nuclear matter \cite{TBF}.
In this work, we follow the somewhat simplified approach originally 
proposed in Ref. \cite{LagPan}, in which the main effect of the three-body force is taken into 
account through a density dependent modification of the intermediate range part of 
$v_{ij}$. Moreover, in view of the weak model dependence of $\eta T^2$ 
(see Fig. \ref{etaT2_1}),  
the full $v_{18}$ potential is replaced by its reduced form $v^\prime_{8}$, and the 
contribution of the non static components is disregarded \cite{shannon}.

In order to obtain $v_{{\rm eff}}$ from Eq.(\ref{def:veff}) the 
expectation value of $H$ in the correlated ground state is evaluated at the two-body 
level of the cluster expansion \cite{shannon}.
The resulting effective interaction reads
\bea
\nonumber
\label{veff:2B}
v_{{\rm eff}} & = & \sum_{i < j} f_{ij}^\dagger \left[ -\frac{1}{m} (\nabla^2 f_{ij}) \right. \\
&  &  \ \ \ \ \ \ \ \ \ \ \ \ \ \ 
\left. - \frac{2}{m} (\bm{\nabla} f_{ij}) \cdot \bm{\nabla} + v_{ij}f_{ij} \right]  \ .
\eea

The radial functions $f^n(r_{ij})$ of Eq.(\ref{def:corrf}) are solutions of a set of 
Euler-Lagrange equations satisfying the boundary conditions $f^1(r_{ij} \geq d) = 1$, 
$f^n(r_{ij} \geq d) = 0$, for $n =2, 3$ and $4$, and  $f^n(r_{ij} \geq d_t) = 0$, 
for $i = 5, 6$ (see, e.g., Ref. \cite{LagPan}). 
%We have used the correlation ranges 
%$d$ and $d_t$ determined in Ref. \cite{APR1} from the minimization of the ground state
%expectation value of an hamiltonian including the Argonne $v_{18}$ NN potential 
%and the Urbana IX three-body potential \cite{TBF}. 

\begin{figure}[htb]
%\vspace*{1.in}
\centerline
{\epsfig{figure=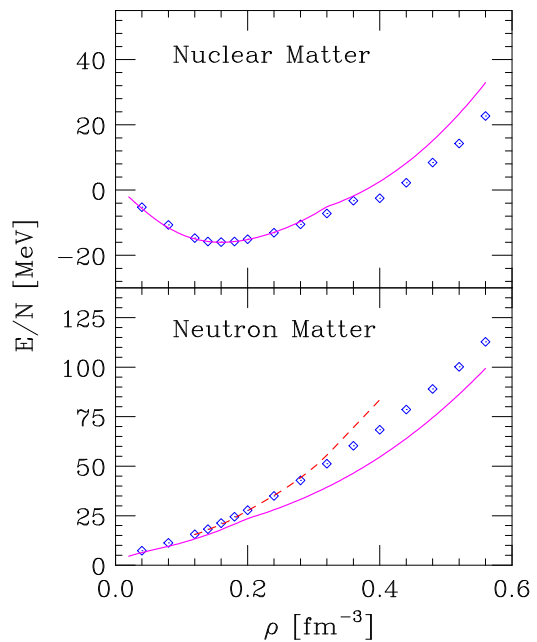,angle=000,width=7cm}}
%\vspace*{.2in}
%\vspace*{.2in}
\caption{ (Color online) 
Energy per particle of symmetric nuclear matter (upper panel) and pure neutron matter
(lower panel). The diamonds represent the results obtained using the effective interaction
discussed in the text in first order perturbation theory with the FG basis,
whereas the solid lines correspond to the results of Akmal Pandharipande and 
Ravenhall \cite{APR}. The dashed line of the lower panel represents the results
of the AFDMC approach or Ref. \cite{AFDMC}. \label{energy_new} }
\end{figure}

The effective interaction of Eq.(\ref{veff:2B}) has been tested computing the energy 
per particle of symmetric nuclear matter and pure neutron matter in first order perturbation 
theory using the FG basis. 
In Fig. \ref{energy_new} our results are compared to those of Refs. \cite{APR} and \cite{AFDMC}.
The calculations of Ref. \cite{APR} (solid lines) have been carried out using a variational 
approach based on the FHNC-SOC formalism, with a hamiltonian including the Argonne $v_{18}$ NN
potential and the Urbana IX three-body potential \cite{TBF}. The results of Ref. \cite{AFDMC}
(dashed line of the lower panel) have been obtained using the $v_8^\prime$ and the same 
three-body potential within the framework of the Auxiliary Field Diffusion Monte Carlo (AFDMC) 
approach.
The results of Fig. \ref{energy_new} show that the effective interaction 
provides a fairly reasonable description of the EOS. 

Note that our approach does not involve adjustable parameters. The correlation ranges
$d$ and $d_t$ have been taken from Ref. \cite{APR1}, while the parameters entering 
the definition of the three-nucleon interaction (TNI) have been determined by 
the authors of Ref. \cite{LagPan} through a fit of nuclear matter equilibrium 
properties.

%%%%%%%%%%%%%%%%%%%%%%%%%%%%%%%%%%%%%%%%%%%%%%%%%%%%%%%%%%%%%%%%%%%%%%%%%%%%%%%%%%%%%
%%%%%%%%%%%%%%%%%%%%%%%%%%%%%%%%%%%%%%%%%%%%%%%%%%%%%%%%%%%%%%%%%%%%%%%%%%%%%%%%%%%%%

Knowing the effective interaction, the in medium scattering probability can be 
readily obtained from Fermi's golden rule. The corresponding cross section
at momentum transfer ${\bf q}$ reads
\beq
\frac{d\sigma}{d\Omega} = \frac{{m^\star}^2}{16 \pi^2} 
\ |\hat{v}_{eff}({\bf q})|^2 \ ,
\label{sigma:medium}
\eeq
$\hat{v}_{eff}$ being the Fourier transform of the effective potential.
The effective mass can also be extracted from the quasiparticle energies computed
in Hartree-Fock approximation.

\begin{figure}[htb]
%\vspace*{1.in}
\centerline
{\epsfig{figure=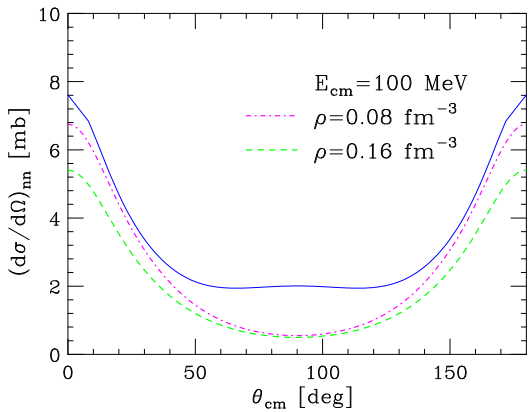,angle=000,width=7.0cm}}
%\vspace*{.2in}
%\vspace*{.2in}
\caption{ (Color online)
Differential neutron-neutron scattering cross section at $E_{cm}=100$ MeV, as a function 
of the scattering angle in the center of mass frame. Solid line: cross section in vacuum, 
calculated with the $v_8^\prime$ potential. Dot-dash line: medium modified cross section 
obtained from the effective interaction described in the text at $\rho = 0.08$ fm$^{-3}$. 
Dashed line: same as the dot-dash line, but for $\rho = 0.16$ fm$^{-3}$.
\label{dsigma} }
\end{figure}

In Fig. \ref{dsigma} the in-medium neutron-neutron cross section at $E_{cm}=100$ MeV 
obtained from the effective potential, with $\rho = \rho_0$ and $\rho_0/2$, is compared 
to the corresponding free space result. As expected, screening of the bare 
interaction leads to an appreciable suppression of the scattering cross section.
%%%%%%%%%%%%%%%%%%%%%%%%%%%%%%%%%%%%%%%%%%%%%%%%%%%%%%%%%%%%%%%%%%%%%%%%%%%%%%%%%%%%%
%%%%%%%%%%%%%%%%%%%%%%%%%%%%%%%%%%%%%%%%%%%%%%%%%%%%%%%%%%%%%%%%%%%%%%%%%%%%%%%%%%%%%

Replacing the cross section in vacuum with the one defined in Eq.(\ref{sigma:medium}),
 the medium modified scattering probability can be readily obtained from Eq.(\ref{Wfree}).
The resulting $W(\theta,\phi)$ can then be used to calculate $\eta T^2$ from 
Eqs.(\ref{eta_AK})-(\ref{eta_sb}).

\begin{figure}[htb]
%\vspace*{1.in}
\centerline
{\epsfig{figure=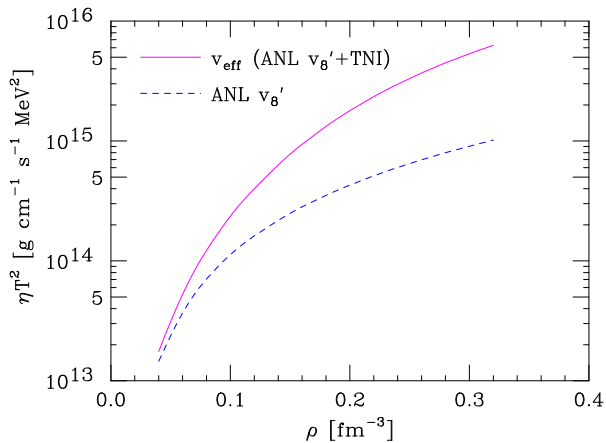,angle=000,width=8.cm}}
%\vspace*{.2in}
%\vspace*{.2in}
\caption{ (Color online)
Solid line: density dependence of $\eta T^2$ computed using the effective interaction 
described in the text. Dashed line: $\eta T^2$ obtained from the free space cross section
corresponding to the $v_8^\prime$ potential.  \label{etaT2_2} }
\end{figure}

The effect of using the medium modified cross section is illustrated in Fig. \ref{etaT2_2}.
Comparison between the solid and dashed lines shows that inclusion of medium 
modifications leads to a large increase of the viscosity, ranging between $\sim$ 75\%
at half nuclear matter density to a factor of $\sim$ 6
 at $\rho = 2 \rho_0$. Such an increase is likely to produce appreciable effects on 
the damping of neutron star oscillations.

%%%%%%%%%%%%%%%%%%%%%%%%%%%%%%%%%%%%%%%%%%%%%%%%%%%%%%%%%%%%%%%%%%%%%%%%%%%%%%%%%%%%%
%%%%%%%%%%%%%%%%%%%%%%%%%%%%%%%%%%%%%%%%%%%%%%%%%%%%%%%%%%%%%%%%%%%%%%%%%%%%%%%%%%%%%

In conclusion, we have computed the shear viscosity of pure neutron matter using
an effective interaction derived from a dynamical model that can also be used to obtain 
the EOS. While our results
are interesting in their own right, as they can be employed in a quantitative analysis 
of the effect of viscosity on neutron stars oscillations, we emphasize
that the work described in this paper should be seen as a first step towards the 
development of a general approach, allowing for a 
consistent calculation of the properties of neutron star matter.

The authors are grateful to V. Ferrari, for drawing their attention to the subject of this
paper, and to R. Schiavilla, for providing a code for the calculation of the NN
scattering cros section. Useful discussions with I. Bombaci
are also gratefully acknowledged.

%%%%%%%%%%%%%%%%%%%%%%%%%%%%%%%%%%%%%%%%%%%%%%%%%%%%%%%%%%%%%%%%%%%%%%%%%

%%%%%%%%%%%%%%%%%%%%%%%%%%%%%%%%%%%%%%%%%%%%%%%%%%%%%%%%%%%%%%%%%%%%%%%%%
\end{document}